\documentclass[prl,twocolumn]{revtex4}

\usepackage{graphicx}
\usepackage{dcolumn}
\usepackage{bm}
\usepackage{amsfonts,amsmath,amssymb}

\newcommand{\cN}{{\cal N}}

\newcommand{\uM}{\underline{M}}
\newcommand{\uN}{\underline{N}}
\newcommand{\uP}{\underline{P}}
\newcommand{\uQ}{\underline{Q}}

\allowdisplaybreaks

\begin{document}

\vspace{-2cm}

\title{Class $\mathcal{S}$ superconformal indices from maximal supergravity}

\author{Ritabrata Bhattacharya$^1$, Abhay Katyal$^{1}$ and Oscar Varela$^{1,2,3}$}

\affiliation{$^1$Department of Physics, Utah State University, Logan, UT 84322, USA
\\
$^2$Black Hole Initiative, Harvard University, Cambridge, MA 02138, USA
\\
$^3$Instituto de F\'\i sica Te\'orica UAM-CSIC, Madrid, 28049, Spain }

\begin{abstract}

\noindent We present a new gauging of maximal supergravity in five spacetime dimensions with gauge group containing ISO(5), involving the local scaling symmetry of the metric, and admitting a supersymmetric anti-de Sitter vacuum. We show this maximal supergravity to arise by consistent truncation of M-theory on the (non-spherical, non-parallelisable) six-dimensional geometry associated to a stack of $N$ M5-branes wrapped on a smooth Riemann surface. The existence of this truncation allows us to holographically determine  the complete, universal spectrum of light operators of the dual four-dimensional $\mathcal{N}=2$ theory of class $\mathcal{S}$. We then compute holographically the superconformal index of the dual field theory at large-$N$, finding perfect agreement with previously known field theory results in specific limits.

\vspace{-3pt}
\begin{center}
{\it Dedicated to Jerome Gauntlett on his 60th birthday.}
\vspace{-15pt}
\end{center}

\end{abstract}


\maketitle

\noindent \textbf{Introduction.} The characterisation of the superconformal field theory (SCFT) that resides on the M5-brane worldvolume is an outstanding open problem in string theory. More tractable limits of this theory have thus received considerable attention over the years. For instance, wrapping a stack of $N$ M5-branes around a punctured Riemann surface $\Sigma$ with a suitable topological twist \cite{Witten:1988ze,Bershadsky:1995qy}, leads the worldvolume theory to an infrared fixed point described by a 4d ${\cal N}=2$ SCFT of so-called class ${\cal S}$ \cite{Gaiotto:2009we}. The field theories in this class are naturally strongly coupled and typically lack Lagrangian descriptions. Useful physical information about them can nevertheless be extracted using dualities. For example, the superconformal index \cite{Kinney:2005ej}, an observable closely related to the partition function, has remarkably been computed in various limits through 4d/2d dualities for class ${\cal S}$ SCFTs \cite{Gadde:2011uv}.

These field theories enjoy known dual gravitational descriptions through the anti-de Sitter/conformal field theory (AdS/CFT) correspondence \cite{Maldacena:2000mw,Lin:2004nb,Gaiotto:2009gz}. The simplest such construction is provided by the half-maximally supersymmetric, ${\cal N}=4$ \footnote{We respectively denote half-maximal supersymmetry by ${\cal N} = 4$ and ${\cal N} = 2$ in 5d supergravity and 4d field theory.}, solution of $D=11$ supergravity presented in \cite{Maldacena:2000mw}. This is of the form $\textrm{AdS}_5 \times (\Sigma \rtimes S^4)$, where the product is warped and $\rtimes$ signifies that the four-sphere $S^4$ is nontrivially fibred over a smooth, unpunctured, constant curvature Riemann surface $\Sigma$ of genus $g \geq 2$. This construction has been generalised along different directions in  \cite{Bah:2011vv,Ferrero:2020laf,Bah:2021mzw}. The dual SCFT arises \cite{Gaiotto:2009gz} as a combination of building blocks of \cite{Gaiotto:2009we}. A specific limit of the index for this SCFT at large-$N$ has been explicitly computed from field theory \cite{Gadde:2011uv}, thus inviting a supergravity calculation that has not been attempted until now.

In this note, we take up the gauntlet. We first make a surprising new observation: $D=11$ supergravity admits a consistent truncation on $\Sigma \rtimes S^4$ to maximal, ${\cal N}=8$, gauged supergravity in $D=5$ spacetime dimensions. The $\textrm{AdS}_5 \times (\Sigma \rtimes S^4)$ solution of \cite{Maldacena:2000mw} arises in this language as a supersymmetry-breaking vacuum. Our  theory contains as a subsector the $D=5$ $\cN=4$ supergravity of \cite{MatthewCheung:2019ehr,Cassani:2019vcl}, and is remarkable in its own right on various accounts. Firstly, all previously known truncations to maximal gauged supergravity occur on spheres ({\it e.g.}~\cite{Nastase:1999cb,Lee:2014mla,Guarino:2015jca}), hyperboloids ({\it e.g.}~\cite{Hohm:2014qga}) or products thereof ({\it e.g.}~\cite{Inverso:2016eet}). Our construction provides the first truncation to maximal gauged supergravity on a distinct non-spherical or hyperbolic background. Secondly, our supergravity contains so-called trombone gaugings, namely, gaugings of the local scaling symmetry of the metric. Our model is the first to provide a concrete example of a higher-dimensional origin for such gaugings with AdS vacua. Thirdly, ours is also the first example of a maximal truncation on a non-parallelisable generalised geometry in the sense of \cite{Lee:2014mla}.

The  existence of this maximal truncation severely constrains the spectrum of Kaluza-Klein (KK) fluctuations of $D=11$ supergravity about the $\textrm{AdS}_5 \times (\Sigma \rtimes S^4)$ solution of \cite{Maldacena:2000mw}. This gives us leverage to compute it using recently introduced spectral techniques \cite{Malek:2019eaz,Malek:2020yue,Varela:2020wty,Cesaro:2020soq}. Finally,  identifying the contributions from short multiplets, we successfully reproduce the large-$N$ superconformal index of \cite{Gadde:2011uv}. This provides the first holographic match of the superconformal index of a class ${\cal S}$ SCFT.

\vspace{2pt}

\noindent \textbf{A new $D=5$ ${\cal N}=8$ supergravity.} Five-dimensional maximal ungauged supergravity has global symmetry $\mathbb{R}^+ \times \mathrm{E}_{6(6)}$. This theory can be gauged by promoting a subset of the vector fields to non-Abelian gauge fields. The gauged theory contains further terms, including covariant derivatives for the 42 $\mathrm{E}_{6(6)}/\mathrm{USp}(8)$ scalars, fermion mass terms or scalar self-interactions.

A $D=5$ $\cN=8$ gauged supergravity is completely specified, including its fermionic sector, by the so-called embedding tensor,
\begin{eqnarray} \label{eq:defX}
X_{\uM \uN}{}^{\uP} \hspace{-1pt} \equiv  \hspace{-1pt}
\Theta_{\uM}{}^\alpha  \hspace{-1pt} (t_\alpha)_{\uN}{}^{\uP}
 \hspace{-1pt} +  \hspace{-2pt} \Big( \hspace{-1pt} \tfrac92 \hspace{-1pt} (t^\alpha)_{\uM}{}^{\uQ}(t_\alpha)_{\uN}{}^{\uP} \hspace{-1pt} - \hspace{-1pt} \delta_{\uM}^{\uQ}\delta_{\uN}^{\uP} \hspace{-0.5pt} \hspace{-1pt} \Big)
\vartheta_{\uQ} ,  \; 
\end{eqnarray}
a constant quantity that determines the gauge group as a subgroup of $\mathbb{R}^+ \times \mathrm{E}_{6(6)}$, and which specifies all couplings induced by the gauging. In (\ref{eq:defX}), $t_\alpha$ are the $\mathrm{E}_{6(6)}$ generators, and $\uM$ and $\alpha$ are fundamental and adjoint indices of $\mathrm{E}_{6(6)}$. Linear constraints imposed by supersymmetry restrict $\Theta_{\uM}{}^\alpha$ and $\vartheta_{\uM}$ to lie in the $\bm{351}$ and $\bm{27}$ representations of $\mathrm{E}_{6(6)}$. Quadratic constraints must also be imposed in order to ensure the closure of the generators (\ref{eq:defX}) into a gauge Lie algebra,
\begin{equation} \label{eq:LieAlg}
[ X_{\uM} , X_{\uN}] = - X_{\uM \uN}{}^{\uP} \, X_{\uP} \; . 
\end{equation}

Theories with non-vanishing $\vartheta_{\uM} $ are popularly known as trombone gaugings. Per (\ref{eq:defX}), these promote the local scaling symmetry of the metric, {\it i.e.}~the $\mathbb{R}^+$ factor of $\mathbb{R}^+ \times \mathrm{E}_{6(6)}$, to a local symmetry. For that reason, these supergravities do not have a Lagrangian and must be described at the level of the equations of motion. While these supergravities have received considerably less attention than their tromboneless counterparts \cite{deWit:2004nw}, their basic details have nevertheless been laid down in \cite{LeDiffon:2008sh}. The bosonic equations of motion can be found in {\it e.g.}~\cite{Blair:2024ofc}.

In order to completely determine our new gauged supergravity, we must specify $\Theta_{\uM}{}^\alpha$ and $\vartheta_{\uM} $. We start by selecting a gauging contained in the maximal subgroup $\textrm{GL}(2) \times \textrm{SL}(6)$ of $\mathbb{R}^+ \times \mathrm{E}_{6(6)}$, and accordingly branch all relevant representations as in (6.6) of \cite{deWit:2004nw}. Then, we choose to couple the gauge fields to the E$_{6(6)}$ generators via $\Theta_{\uM}{}^\alpha$ charges split as
\begin{equation} \label{eq:TableGaugings}
\begin{tabular}{c| c c c c} 
~&~&~\\[-4mm]
&$({\bf 1},{\bf 15})$
&&&$({\bf 2},\overline{\bf 6})$
\\ \hline
~&~&~\\[-3.5mm]
$({\bf 1},{\bf 35})$
&$({\bf 1},{\bf 21}) + ({\bf 1},{\bf 105}) $
&&&$({\bf 2},\overline{\bf 6})+({\bf 2},\overline{\bf 84}) $
\\
$({\bf 3},{\bf 1})$
& --- 
&&&$({\bf 2},\overline{\bf 6})$
\\
$({\bf 2},{\bf 20})$
&$({\bf 2},\overline{\bf 6})+({\bf 2},\overline{\bf 84}) $
&&&$ ({\bf 1},{\bf 105}) $
\\ \hline
\end{tabular}
\end{equation}
and to $\mathbb{R}^+$ by a trombone $\vartheta_{\uM} $ in the $({\bf 2}, \overline{\bf 6})$. This is a very rich gauging, where all allowed components of the embedding tensor, except the $({\bf 3},{\bf 15})$, are turned on (see (6.7) of \cite{deWit:2004nw}). Now, we introduce fundamental indices of SL(2), $x=1,2$, and SL(6), $A=1, \ldots , 6$, to denote the various components of $\Theta_{\uM}{}^\alpha$ and $\vartheta_{\uM} $ above by
\begin{eqnarray} \label{eq:EmTenComps}
& ({\bf 1},{\bf 21}) : \theta_{AB} \equiv \theta_{(AB)} ,  & ({\bf 1},{\bf 105}) : \theta_{AB}{}^C{}_D \equiv \theta_{[AB}{}^C{}_{D]} , \nonumber  \\[5pt]
& ({\bf 2}, \overline{\bf 6}) : \xi^{xA} \sim \vartheta^{xA} , &  ({\bf 2}, \overline{\bf 84}) : \xi^{xAB}{}_C \equiv \xi^{x[AB]}{}_C . \qquad 
\end{eqnarray}
We identify the $({\bf 2}, \overline{\bf 6})$ components of  $\Theta_{\uM}{}^\alpha$ and $\vartheta_{\uM} $, thereby evading the argument \cite{deWit:2004nw} that tromboneless gaugings contained in $\textrm{SL}(2) \times \textrm{SL}(6)$ must involve the $ ({\bf 1},{\bf 21})$ component only. Next, we further break $\textrm{SL}(6) \supset \textrm{SO}(6) \supset \textrm{SO}(5) \supset \textrm{SU}(2) \times \textrm{U}(1)$ and introduce SO(5), SO(2)$\sim$U(1) and SO(3)$\sim$SU(2) indices $i = 1, \ldots , 5$, $a=1,2$, $\alpha=3,4,5$, so that $A=(i,6)$ and $i=(a,\alpha)$. Finally, we take the non-vanishing components in (\ref{eq:EmTenComps}) to be
\begin{eqnarray} \label{eq:EmTenCompsExpl}
 \theta_{ij} \equiv g_1 \delta_{ij}  \; , \qquad  \theta_{\alpha\beta}{}^6{}_\gamma = -6  \, \kappa \,  g_2  g_3 g_1^{-1}  \,  \epsilon_{\alpha\beta\gamma} , \nonumber \\*
\xi^{16} = -\tfrac{3}{8} \, \vartheta^{16}  = -\tfrac{1}{8}   g_2  \sqrt{-\kappa}  \; , \quad \xi^{2ab}{}_6 = - g_3 \sqrt{-\kappa} \, \epsilon^{ab}  \; . 
\end{eqnarray}
Here, $g_1$,  $g_2$ and $g_3$ are coupling constants, $\kappa$ is either $\kappa=0$ or $\kappa=-1$, and $\delta_{ij}$, $\epsilon_{ab}$, $\epsilon_{\alpha\beta\gamma}$ are the usual invariant tensors of SO(5), U(1) and SU(2). 

By construction, the embedding tensor (\ref{eq:defX}) with (\ref{eq:EmTenComps}), (\ref{eq:EmTenCompsExpl}) satisfies the linear constraints. We have also verified the quadratic constraint (\ref{eq:LieAlg}). It is apparent from (\ref{eq:EmTenCompsExpl}) that the gauge group must contain at least $\textrm{SU}(2) \times \textrm{U}(1)$, but less so that, in fact, it contains $\textrm{ISO}(5) \equiv \textrm{SO}(5) \ltimes \mathbb{R}^5$. This is immediate for $\kappa = 0$, as our theory reduces to the conventional ISO(5) gauging contained in the $\textrm{CSO}(p,q,r)$ family of  \cite{Andrianopoli:2000fi,Gunaydin:1985cu}. For $\kappa = -1$, our supergravity is not in the $\textrm{CSO}(p,q,r)$ class, and some work is required to show that the gauge group is $B_2 \ltimes \textrm{ISO}(5)$. Here, $B_2$ is the two-dimensional Borel subgroup of GL(2), acting on the $\mathbb{R}^5$ translations of ISO(5). This theory admits a formulation with only 17 gauge fields and 10 two-forms.

It is useful to consider smaller sectors. As explained in \cite{Blair:2024ofc}, even if the parent maximal supergravity gauges the trombone, its subsectors may or may not involve such gaugings from their point of view. Two interesting sectors are attained as follows. The first retains singlets under the U(1) subgroup of $\textrm{E}_{6(6)}$ defined in (5.20) of \cite{Cassani:2019vcl}. We have verified that this truncates our theory to the tromboneless $D=5$ $\cN=4$ supergravity of \cite{MatthewCheung:2019ehr,Cassani:2019vcl}. Interestingly, this U(1) is not contained in the gauge group $B_2 \ltimes \textrm{ISO}(5)$, and is an example of a type of $\cN=4$ truncations of $\cN=8$ supergravity described in \cite{Guarino:2024gke}. A second interesting subsector does retain singlets under the $\textrm{SU}(2) \times \textrm{U}(1) \subset \textrm{SO}(5)$ subgroup of the gauge group. The resulting theory is non-Lagrangian, non-supersymmetric, and its scalar sector includes three dilatons, $\varphi_0$, $\varphi_1$, $\varphi_2$, and three axions that parametrise $\textrm{GL}(3)/\textrm{SO}(3)$. 

Both subsectors overlap in a Lagrangian theory with a SO$(1,1)^2$ scalar manifold obtained by setting $\varphi_2$ and the axions to zero. In our conventions, this model reads
\begin{equation} \label{eq:littlemodel}
e^{-1} {\cal L} = R -6 (\partial \varphi_0)^2 -6 (\partial \varphi_1)^2 -V \; ,
\end{equation}
and has scalar potential
{\setlength\arraycolsep{1pt}
\begin{eqnarray} \label{eq:littlemodelPot}
V &=& \tfrac{1}{2} \, g_2^{4} \, g_1^{-2} \, \kappa^2 \, e^{-4\varphi_0 + 8 \varphi_1} -2 g_2^2 \, \kappa \, e^{-4\varphi_0 + 2 \varphi_1} \\[1.6pt] 
\nonumber 
&& -\tfrac32 g_1^2 \big( 4 e^{-2\varphi_0 }+  e^{4\varphi_1 }  \big) \;  ,
\end{eqnarray}
}where we have set $g_2 = g_3$ for simplicity \footnote{This model matches (3.7), (3.8) of \cite{MatthewCheung:2019ehr}, with $\varphi_{0 \textrm{here}} = \lambda_{\textrm{there}} + 2 \phi_{\textrm{there}}$, $\varphi_{0 \textrm{here}} = 2\lambda_{\textrm{there}} -  \phi_{\textrm{there}}$, $\kappa_{\textrm{here}} = l_{\textrm{there}}$, $g_{1 \textrm{here}} = g_{2 \textrm{here}} = g_{\textrm{there}}$ and everything else set to zero.}. For $\kappa = -1$, the model (\ref{eq:littlemodel}), (\ref{eq:littlemodelPot}) admits an AdS vacuum located at
\begin{equation} \label{eq:AdSLoc}
e^{6 \varphi_0 } = 2 \, g_2^4 \, g_1^{-4} \; , \qquad e^{6 \varphi_1 } = 2 \, g_2^2 \, g_1^{-2} \; ,
\end{equation} 
with squared radius $L^2= 2^{4/3} g_1^{-10/3} g_2^{4/3} $. This solution preserves $\cN=4$ supersymmetry and, by construction, $\textrm{SU}(2) \times \textrm{U}(1)$ within the full $D=5$ $\cN=8$ supergravity. We have computed the mass spectrum of this vacuum within our maximal supergravity and have allocated the corresponding mass states in supermultiplets of the 4d $\cN=2$ superconformal algebra, $\mathfrak{su}(2,2|2)$. We find \footnote{Our notation for the $\mathfrak{su}(2,2|2)$ supermultiplets is as in \cite{Cordova:2016emh}, with states denoted $[j_1;j_2]_\Delta^{(\ell;n)}$, where $j_ {\textrm{there}}= 2 j_{1\textrm{here}} $, $\bar{j}_ {\textrm{there}}= 2 j_{2\textrm{here}} $, $R_ {\textrm{there}}= 2 \ell_{\textrm{here}} $, $r_ {\textrm{there}}= 2 n_{\textrm{here}} $.}
\begin{eqnarray} \label{eq:KK0Spec}
& A_2\bar{A}_2[0;0]_2^{(0;0)} \oplus B_1\bar{B}_1[0;0]_4^{(2;0)} \nonumber \\
& \oplus   \, A_2\bar{B}_1 [0;0]_3^{(1;1)} \,  \oplus B_1\bar{A}_2  [0;0]_3^{(1;-1)}  \; .
\end{eqnarray}
The first line here reproduces the $D=5$ $\cN=4$ result of \cite{MatthewCheung:2019ehr}. The $A_2\bar{A}_2$ multiplet contains the massless graviton and the $\textrm{SU}(2) \times \textrm{U}(1)$ massless gauge fields.

\vspace{2pt}

\noindent \textbf{M-theory uplift.} Our $D =5$ supergravity originates by truncation of $D=11$ supergravity on the internal manifold $\Sigma \rtimes S^4$ of the 16-supercharge solution of \cite{Maldacena:2000mw}. We will show this using exceptional generalised geometry (EGG) \cite{Coimbra:2011ky}. In this formalism, the field content of $D=11$ supergravity is repacked in a generalised vielbein $\hat{E}_{\underline{M}}{}^M$. Both indices, local, $M$, and, global, $\underline{M}$, are in the fundamental of $\textrm{E}_{6(6)}$. The associated torsion, $X_{\uM \uN}{}^{\uP}$, is defined as
\begin{equation}
L_{\hat{E}_{\uM}} \hat{E}_{\underline{N}} = X_{\underline{M}\underline{N}}{}^{\underline{P}} \, \hat{E}_{\underline{P}} \; , 
\end{equation}
where $L$ is the generalised Lie derivative \cite{Coimbra:2011ky} and we omit local indices. A consistent truncation to maximal supergravity exists iff the torsion  is constant \cite{Lee:2014mla,Hohm:2014qga}, thereby giving rise to the lower-dimensional embedding tensor.

The solution of \cite{Maldacena:2000mw} has been described within EGG \cite{Cassani:2019vcl}, and we build on their results. The branching (5.11) of \cite{Cassani:2019vcl} leads us to consider a generalised frame
\begin{equation} \label{eq:frameblocks}
\hat{E}_{\uM} = \big( \hat{E}_x , \, \hat{E}^{x i}  , \, \hat{E}_i , \,  \hat{E}_{ij} \big) \; , 
\end{equation}
with indices $x=1,2$, $i=1, \ldots , 5$ as above. Each entry in (\ref{eq:frameblocks}) is a generalised vector of the form $V = (v , \omega, \sigma)$, with $v$ a vector, $\omega$ a two-form and $\sigma$ a five-form on $\Sigma \rtimes S^4$. To specify these, we introduce vielbein one-forms $e^x$ and their dual vectors $\hat{e}_x$ on $\Sigma$. We also introduce functions $y^i (\theta, \psi , \chi , \phi )$ of the $S^4$ angles constrained as $\delta_{ij} y^i y^j = 1$, and the $S^4$ Killing vectors $v^{ij}$. We then take
\begin{eqnarray} \label{eq:DirProdFrame}
&& \hat{E}_x = e^{\Upsilon} \, ( \hat{e}_x ,\;  0  , \; 0) \; , \nonumber \\[5pt]
&& \hat{E}^{ x i}   = e^{\Upsilon} \,  ( 0, \; R\, e^x \wedge d y^i,  \; e^x \wedge (- y^i \textrm{vol}_{4} + R\, d y^i \wedge A ))   ,  \nonumber \\[5pt]
&& \hat{E}_i = e^{\Upsilon} \,  (0,  \; y_i \textrm{vol}_\Sigma , \; \textrm{vol}_\Sigma \wedge (-R*_4 d y_i + y_i A ))   \; ,  \nonumber \\[5pt]
&& \hat{E}_{ij} = e^{\Upsilon} \,  ( v_{ij} , \; R^2*_4\! (d y_i\wedge d y_j) + \iota_{v_{ij}} A  , \;  0)  \; .
\end{eqnarray}
Here, $R$ is constant, $\textrm{vol}_\Sigma$ and $\textrm{vol}_4$ are volume forms, the Hodge dual is taken w.r.t.~the round $S^4$ metric, $\iota$ denotes interior product, and $A$ is the three-form on $S^4$ given in (B.5) of \cite{Cassani:2019vcl}. The fibration is implemented by the group action of the local $\mathbb{R} \oplus \mathfrak{e}_{6(6)}$ Lie algebra element \cite{Cassani:2019vcl}
\begin{equation} \label{eq:FibrElem}
\Upsilon =  -R \, \big(  v_{12} \otimes \upsilon \; , \;  \upsilon \wedge [ R^2*_4\! (d y_1\wedge d y_2) + \iota_{v_{12}} A ] \big) \; ,
\end{equation}
featuring $\upsilon$, the spin connection on $\Sigma$, such that $d e^x = -\epsilon^{x}{}_y \, \upsilon \wedge e^y$ and $d\upsilon = \tfrac{\kappa}{R_\Sigma^2} \, \textrm{vol}_\Sigma$, with $\kappa=0,\pm1$ the constant curvature, $R_{xy} = \tfrac{\kappa}{R_\Sigma^2} \, \delta_{xy}$, of $\Sigma$ and $R_\Sigma$ a constant.

A lenghty calculation shows that the generalised torsion $X_{\uM \uN}{}^{\uP}$ associated to our frame (\ref{eq:frameblocks}) with (\ref{eq:DirProdFrame}), (\ref{eq:FibrElem}) is constant, and indeed matches the embedding tensor (\ref{eq:defX}) with (\ref{eq:EmTenComps}), (\ref{eq:EmTenCompsExpl}), with $g_1 = R^{-1}$, $g_2 = g_3 = R_\Sigma^{-1}$. This shows the consistency of the truncation. Consistency goes through even if our generalised frame does not extend globally in EGG, due to its definition in terms of the frames $e^x$, $\hat{e}_x$ on $\Sigma$. These are only local in conventional geometry,  even if the full solution $\textrm{AdS}_5 \times (\Sigma \rtimes S^4)$ is, of course, globally defined. We thus obtain a maximal consistent truncation on a non-parallelisable, local generalised identity structure. This is similar to the truncations on conventional $G$-structures of \cite{Gauntlett:2007ma}, which need to be defined only locally.

The $D=11$ uplift of our $D=5$ $\cN=8$ supergravity generically induces non-linear deformations on both $\Sigma$ and $S^4$. The U(1)-invariant sector considered above leaves $\Sigma$ intact but deforms $S^4$. Selecting the U(1)-invariant generalised vectors, (5.21) of \cite{Cassani:2019vcl}, out of our frame (\ref{eq:frameblocks}), we have explicitly recovered the truncation of \cite{MatthewCheung:2019ehr,Cassani:2019vcl}. On the other hand, the $\textrm{SU}(2) \times \textrm{U}(1)$-invariant sector deforms both $\Sigma$ and $S^4$, leaving an $S^2 \times S^1$ inside the latter untouched. The uplifted metric for this sector with zero axions and vectors, and $\kappa = -1$, reads \footnote{For $\varphi_2=0$, this agrees with (5.36) of \cite{Cassani:2019vcl} with $R_\Sigma = R$ and $e^{2\varphi_{0 \textrm{here}}} = e^{2\varphi_{1 \textrm{here}}} = \sqrt{2} \, \Sigma^{-1}_{\textrm{there}}$.}
{\setlength\arraycolsep{0pt}\begin{eqnarray} \label{eq:UpliftedMetric}
& ds_{11}^2 = \bar{\Delta}^{\frac13} \Big( ds_5^2 +R_\Sigma^2 \, e^{4 \varphi_0 - 2 \varphi_1 - 2\varphi_2} \, \frac{ dx^2 + e^{4\varphi_2} dy^2}{y^2}  \quad \\
& +\hspace{-1pt} R^2 \hspace{-1pt} \big[ \hspace{-1pt} e^{2 \varphi_0} d\theta^2 \hspace{-2pt} + \hspace{-2pt}  \frac{e^{-4 \varphi_0}}{\bar{\Delta}} \hspace{-1pt} \cos^2 \hspace{-2pt} \theta ds^2(S^2) \hspace{-2pt}  + \hspace{-2pt} \frac{e^{-2 \varphi_0+4\varphi_1}}{\bar{\Delta}} \hspace{-1pt} \sin^2 \hspace{-2pt} \theta \hspace{-0.5pt} ( \hspace{-1pt} D\psi \hspace{-1pt})^2  \hspace{-1pt} \big] \hspace{-2pt} \Big) \nonumber
\end{eqnarray}
}where $\bar{\Delta} \hspace{-1pt} = \hspace{-2pt}  e^{-4 \varphi_0+4\varphi_1} \hspace{-1pt} \cos^2 \hspace{-1pt}  \theta \hspace{-1pt}  + \hspace{-1pt}  e^{-6 \varphi_0} \hspace{-1pt} \sin^2 \hspace{-1pt} \theta$ and $D\psi = d\psi + \upsilon$, with $\upsilon = -dx/y$. The AdS vacuum (\ref{eq:AdSLoc}) uplifts via these formulae to the half-maximal $D=11$ solution of \cite{Maldacena:2000mw}. 

\vspace{2pt}

\noindent \textbf{Universal spectrum.} We will now compute the complete,  universal KK spectrum of this solution, valid for any smooth, compact, genus $g \geq 2$ Riemann surface $\Sigma$. Our maximal truncation fixes the algebraic structure of the spectrum, with the gravitons playing a leading role.

We thus start by computing the KK graviton spectrum. This requires solving the eigenvalue problem \cite{Bachas:2011xa}
\begin{equation} \label{eq:pde_main}
\bar{\Delta}^{-\frac32} \, \bar{g}^{-\frac12} \, \partial_m \big(\bar{\Delta}^{\frac32} \, \bar{g}^{\frac12} \, \bar{g}^{mn}\partial_n \big) {\cal Y} =-M^2L^2 {\cal Y} \; .
\end{equation}
Here, we have restored local indices $m=1,\dots,6$, and the metric, $\bar{g}_{mn}$, and its determinant, $\bar{g}$, are those in (\ref{eq:UpliftedMetric}) evaluated in the vacuum (\ref{eq:AdSLoc}). The ansatz
\begin{equation} \label{eq:EigenFunSep}
{\cal Y} = u^{\ell/2} (1-u)^{|n|/2} H(u) \,  \varphi( x,y) \, e^{in\psi} \, Y_\ell^m(\chi ,  \phi) \; ,
\end{equation}
with $\ell$ and $n$ the $\textrm{SU}(2) \times \textrm{U}(1)$ R-symmetry quantum numbers , $Y_\ell^m$ the $S^2$ spherical harmonics and $u \equiv \cos^2 \theta$, separates (\ref{eq:pde_main}) into a hypergeometric  \footnote{The parameters are $a =\tfrac14 (2\ell +2 |n| +3)  + A$, $b =\tfrac14 (2\ell +2 |n| +3)  - A$ and $c = \ell + \tfrac32$, with 
$A \equiv \tfrac14  \sqrt{ 2L^2 M^2 +4E-4\ell(\ell+1)-2n^2+9 }$, 
where $E$ is the separation constant.},
\begin{equation} \label{eq:Hypergeom}
u(1-u)H''(u) +(c-(a+b+1)u)H'(u) -abH(u) =0    ,
\end{equation}
and the eigenvalue problem, $D_n \varphi (x,y) = E \, \varphi (x,y)$, for the Maass Laplacian on $\Sigma$,
\begin{equation} \label{eq:MaassLap}
D_n  \equiv y^2 \big( \tfrac{\partial^2}{\partial x^2}  + \tfrac{\partial^2}{\partial y^2}  \big) +2 i n y \tfrac{\partial}{\partial x} \; .
\end{equation}
The latter is well-known from the theory of automorphic forms, and has already appeared in physics applications: see \cite{DHoker:2022dxx}. The eigenvalues of $D_n$ on $\Sigma$ are discrete, and those universally present for all $\Sigma$ are \cite{Elstrodt}  \footnote{In addition, $D_n$ may have other discrete, non-universal ({\it i.e.} specific to each $\Sigma$) eigenvalues, which we disregard.}
\begin{equation} \label{eq:MaassEigenv}
E_{nj} = ( |n| - j)( |n| - j - 1) \; , \quad \textrm{for }  |n| \geq  1 \; ,
\end{equation}
and have the genus-dependent multiplicities \cite{Shimura}
\begin{equation} \label{eq:Maassmult}
m_{nj} = (g-1) ( 2 |n| - 2j - 1) \; .
\end{equation}
The range of the quantum number $j$ is \cite{Elstrodt}
\begin{equation} \label{eq:MaassRanges}
j = 0 , \ldots ,  |n|-1 \; , \quad \textrm{for fixed  }  |n| \geq  1 \; ,
\end{equation}
bounded by the U(1) charge $n$ due to the fibration. Turning to equation (\ref{eq:Hypergeom}), regularity of the eigenfunctions forces $a$ to be a non-negative integer, $a \equiv -\tilde{\jmath}$. Defining $k \equiv 2 \tilde{\jmath} + \ell + |n|$, we finally get masses
\begin{equation} \label{eq:BE_mass}
L^2 M^2_{k\ell nj} = 2k(k+3) + 2 \ell(\ell+1) + n^2 -2 E_{nj} \; ,
\end{equation}
with quantum numbers ranging as (\ref{eq:MaassRanges}) and 
\begin{eqnarray} \label{eq:QNRanges}
& k=0 , 1 , \ldots \; , \quad 
\ell = 0, 1, \ldots , k \; , \\
& n = -(k-\ell) , -(k-\ell-2 ) , \ldots , (k-\ell-2 ) , (k-\ell) \; . \nonumber
\end{eqnarray}
Further analysis allows us to characterise the set of universal graviton eigenstates ${\cal Y}^\Lambda$ at each KK level $k$. These correspond to $(g-1)$ copies of the $S^4$ spherical harmonics, in the $(0,k)$ irrep of SO(5), branched under $\textrm{SU}(2) \times \textrm{U}(1)$, and augmented with $(2\ell+1) \, m_{nj}$ additional eigenstates of $D_n$ for all possible $\ell$, $n$, $j$ allowed by (\ref{eq:QNRanges}), (\ref{eq:MaassRanges}). 

In order to compute the rest of the spectrum, we employ the methods of \cite{Malek:2019eaz,Malek:2020yue,Varela:2020wty,Cesaro:2020soq}. We will only outline this long calculation, leaving further details to be presented elsewhere. The linearised gauged supergravity can be interpreted as level $k=0$ in the KK expansion. All states of spin $0 \leq s \leq 2$ at $k \geq 1$ arise by tensoring the $D=5$ $\cN=8$ supergravity multiplet with the graviton eigenstates. This process fixes the multiplicities for all states and allows one to obtain their masses by diagonalising finite-dimensional mass matrices level by level. The latter are obtained by linearising the equations of motion of exceptional field theory \cite{Hohm:2013pua}, using our frame (\ref{eq:frameblocks}) and keeping track of trombone terms. These mass matrices depend on gauged supergravity data and on the constant quantities $({\cal T}_{\uM})_\Lambda{}^\Sigma$, obeying $L_{\hat{E}_{\uM}} \, {\cal Y}_\Lambda  = -({\cal T}_{\uM})_\Lambda{}^\Sigma \, {\cal Y}_\Sigma$ and the algebra (\ref{eq:LieAlg}). The explicit construction of Maass eigenfunctions on compact $\Sigma$ is an open problem, but the above algebraic arguments are enough to determine $({\cal T}_{\uM})_\Lambda{}^\Sigma$. Finally, we manage to allocate the resulting states into supermultiplets of $\mathfrak{su}(2,2|2)$, which provides a reassuring crosscheck on our algebraic prescription. 

Thus we obtain the complete, universal spectrum at KK level $k=0, 1 , \ldots$ It consists of $(g-1)$ copies of
\begin{eqnarray} \label{eq:FullKKSpec}
& A_2\bar{A}_2[0;0]_{2k+2}^{(k;0)} \quad \,  \oplus B_1\bar{B}_1[0;0]_{2k+4}^{(k+2;0)} \nonumber \\
& \oplus \, A_2\bar{B}_1 [0;0]_{2k+3}^{(k+1;1)} \oplus B_1\bar{A}_2  [0;0]_{2k+3}^{(k+1;-1)}   \\
& \bigoplus_{\ell=0}^{k-1}\bigoplus_{n=-(k-\ell)}^{\prime k-\ell}\bigoplus_{j=0}^{|n|-1} | 2 |n|-2j-1 | \, L\bar{L} [0;0]_{E_{k\ell n j}}^{(\ell; n)} \nonumber \; , 
\end{eqnarray}
where the last line is only present for $k   \geq 1$; $j \hspace{-2pt} = \hspace{-2pt}  0$ if $n \hspace{-2pt}  = \hspace{-2pt}  0$; $\oplus^\prime$ denotes that $n$ changes in steps of 2 as in (\ref{eq:QNRanges}); and 
{\setlength\arraycolsep{-10pt}
\begin{eqnarray}
E_{k\ell n j} \equiv \hspace{-4pt} \sqrt{4+2k(k+3)+2\ell(\ell+1)+n^2-2E_{nj}} \qquad  \qquad  \qquad 
\end{eqnarray}
}is the dimension of all multiplets, including the short ones once the shortening conditions \cite{Cordova:2016emh} are considered.

The KK spectrum (\ref{eq:FullKKSpec}) maps to the spectrum of light operators of the dual SCFT. It reproduces the gauged supergravity result (\ref{eq:KK0Spec}) (except for the $(g-1)$ dependence, which is a higher-dimensional effect) and extends it to all KK levels. Our $A_2\bar{A}_2$ tower also matches \cite{Chen:2019ydk}. The scaling of the spectrum with the genus of $\Sigma$, expected from the field theory construction \cite{Gaiotto:2009gz}, arises from the multiplicities (\ref{eq:Maassmult}) of the Maass Laplacian eigenvalues. There are no irreps other than those of $\mathfrak{su}(2,2|2)$ shown in (\ref{eq:FullKKSpec}), consistent with the lack of flavour symmetries of the dual SCFT. Finally, our methods are insensitive to topological modes, but these are immediate to pin down as the $3(g-1)$ complex structure moduli of $\Sigma$, dual to the marginal couplings of the SCFT. By construction, these do not enter the index (see \cite{Kinney:2005ej}), to which we now turn.

\vspace{2pt}

\noindent \textbf{Superconformal index.} The 4d $\cN=2$ SCFT index is given by \cite{Kinney:2005ej,Gadde:2011uv}
\begin{equation} \label{eq:SCI}
\mathcal{I}(\rho,\sigma,\tau)=\text{Tr}(-1)^F\rho^{\frac{1}{2}\delta_{1-}}\sigma^{\frac{1}{2}\delta_{1+}}\tau^{\frac{1}{2}\tilde{\delta}_{2\dot{+}}}e^{-\beta\tilde{\delta}_{1\dot{-}}} \ ,
\end{equation} 
with $\rho$, $\sigma$, $\tau$ superconformal fugacities, $F$ the fermion number, and charges
{\setlength\arraycolsep{0pt}
\begin{eqnarray}
& \tilde{\delta}_{1\dot{-}} \equiv \Delta-2j_2-2\ell+2n , \; \delta_{1-} \equiv\Delta-2j_1-2\ell-2n, \quad  \; \;  \nonumber \\
& \delta_{1+} \equiv \Delta+2j_1-2\ell-2n, \; \tilde{\delta}_{2\dot{+}} \equiv \Delta+2j_2+2\ell+2n . \quad \;\;
\end{eqnarray} 
}The trace is over the Hilbert space on $S^3$, in radial quantisation. The index captures contributions of states with $\tilde{\delta}_{1\dot{-}}=0$, so that ${\cal I}$ does not depend on $\beta$.

\begin{table}
\centering
\begin{tabular}{l c l c l c l}
\text{multiplet} && descendant &&\ state  &&\ contribution\\
\hline 
$A_2\bar{A}_2$ &&\ $\mathcal{Q}_{1+}\tilde{\mathcal{Q}}_{1\dot{+}}$ &&\ $[\frac12 ; \frac12 ]_{2k+3}^{(k+1;0)}$  &&\ $\sigma\tau^{2k+3}$ \\[3pt]
$B_1\bar{B}_1$ &&\ 1 &&\ $[0;0]_{2k+4}^{(k+2;0)}$ &&\ $\tau^{2k+4}$ \\[3pt]
$B_1\bar{A}_2$ &&\ $\mathcal{Q}_{2+}\mathcal{Q}_{2-}$ &&\ $[0;0]_{2k+4}^{(k;-2)}$ &&\ $\rho^4\sigma^4\tau^{2k}$ \\[3pt]
$B_1\bar{A}_2$ &&\ $\tilde{\mathcal{Q}}_{1\dot{+}}\tilde{\mathcal{Q}}_{2\dot{+}}$ &&\ $[0;1]_{2k+4}^{(k+1;0)}$  &&\ $\rho\sigma\tau^{2k+4}$ 
\end{tabular}
\caption{Contributions to the index \label{tab:SCI_contri}} 
\end{table}

Only states in short multiplets may contribute to the index. Writing out the state content of the short multiplets in (\ref{eq:FullKKSpec}) with the help of \cite{Cordova:2016emh}, we identify the contributions summarised in table \ref{tab:SCI_contri}. Taking into account the genus-dependent multiplicity and the overall contributions from derivatives, we obtain the single-letter index
{\setlength\arraycolsep{0pt}
\begin{eqnarray} \label{eq:SCI}
\mathcal{I}(\rho,\sigma,\tau) &=& (g-1) (1-\rho\tau)^{-1} (1-\sigma\tau)^{-1}  
 \\
&& \times \hspace{-3pt} \sum_{k=0}^{\infty}\left(\sigma\tau^{2k+3} +\tau^{2k+4} +\rho^4\sigma^4\tau^{2k}+\rho\sigma\tau^{2k+4} \right) ,  \nonumber
\end{eqnarray}
}or, after summing the series,
\begin{equation} \label{eq:SCISummed}
\mathcal{I}(\rho,\sigma,\tau) = (g-1)\frac{\sigma\tau^{3}+\tau^{4} +\rho^4\sigma^4+\rho\sigma \tau^{4} }{(1-\tau^2)(1-\rho\tau)(1-\sigma\tau)}\ .
\end{equation}

The index was computed in \cite{Gadde:2011uv}, generically at finite $N$, from the SCFT side for various limiting cases of the fugacities. We holographically match their large-$N$ results whenever available, and predict the large-$N$ behaviours otherwise. The Hall-Littlewood index, ${\cal I}_{\textrm{HL}} \equiv {\cal I}(0,0,\tau)$, receives contributions from $B_1\bar{B}_1$ only, and (\ref{eq:SCISummed}) yields
\begin{equation} \label{eq:SCIHL}
{\cal I}_{\textrm{HL}} = (g-1) \, \frac{\tau^{4}}{1-\tau^2} \; , 
\end{equation}
in agreement with the large-$N$ result (5.49) of \cite{Gadde:2011uv} after taking the plethystic exponential (and up to an overall sign). Similarly, (\ref{eq:SCISummed}) provides the large-$N$ limit of the Schur index, ${\cal I}_{\textrm{S}} \equiv {\cal I}(\rho,\sigma,\rho)$, the Macdonald index, ${\cal I}_{\textrm{M}} \equiv {\cal I}(\rho,0,\tau)$, or the Coulomb branch index, ${\cal I}_{\textrm{C}} \equiv {\cal I}(\rho,\sigma,0)$.

\vspace{2pt}

\noindent \textbf{Final remarks.} It would be interesting to compute the large-$N$ limit of these other $\cN=2$ class ${\cal S}$ indices from the finite-$N$ results of \cite{Gadde:2011uv} and compare to our holographic predictions. More generally, our formalism opens up the possibility to compute the indices of other $\cN=1$ SCFTs that similarly arise on M5-branes wrapped on $\Sigma$ \cite{Maldacena:2000mw,Bah:2011vv}. 

Indeed, our supergravity turns out to belong to a larger family of $D=5$ $\cN=8$ gaugings that arise by truncation on the geometries of \cite{Bah:2011vv,Maldacena:2000mw}. Intriguingly, all these truncations go through on $\kappa=1$ Riemann surfaces as well, but the embedding tensors become complex (see (\ref{eq:EmTenCompsExpl})). Further truncating to the $\cN=4$ and $\cN=2$ subsectors of \cite{MatthewCheung:2019ehr,Cassani:2019vcl} and \cite{Cassani:2020cod}, the complex components disappear. 

Further details will be given elsewhere.

\vspace{2pt}

\noindent \textbf{Acknowledgements.} We are pleased to thank I.~Bah, E.~D'Hoker, M.~Gutperle, D.~Jafferis, G.~Larios,  A.~Malmendier, C.~N\'u\~nez, M.~Pico and L.~Rastelli for helpful discussions. This work was supported by NSF grant PHY-2310223 and, partially, by Spanish Government grants CEX2020-001007-S and PID2021-123017NB- I00, funded by MCIN/AEI/10.13039/501100011033 and by ERDF A way of making Europe.

\end{document}